\documentclass[a4paper]{jpconf}
\usepackage{graphicx}
\begin{document}
\title{Evolution of quantum criticality in the system CeNi$_9$Ge$_4$}
\author{H Michor$^1$, D T Adroja$^2$, A D Hillier$^2$, M M Koza$^3$, S Manalo$^1$, \\ 
C Gold$^4$, L Peyker$^4$ and E-W Scheidt$^4$}
\address{$^1$Institut f\"ur Festk\"orperphysik, Technische Universit\"at Wien, A-1040 Wien, Austria}
\address{$^2$ISIS Facility, Rutherford Appleton Laboratory Chilton, Didcot OX11\,0QX, UK}
\address{$^3$Institut Laue–Langevin, B.P. 156, F-38042, Grenoble Cedex 9, France}
\address{$^4$ CPM, Institut f\"{u}r Physik, Universit\"{a}t Augsburg, 86159 Augsburg, Germany}
\ead{michor@ifp.tuwien.ac.at}

\begin{abstract}
The heavy fermion system CeNi$_9$Ge$_4$ exhibits a paramagnetic ground state
with remarkable features such as: 
a record value of the electronic specific heat coefficient in systems with a paramagnetic ground state, 
$\gamma=C/T\simeq 5.5$ J/mol\,K$^2$ at 80\,mK, a temperature-dependent Sommerfeld--Wilson ratio, $R=\chi/\gamma$, 
below 1 K and an approximate single ion scaling of the $4f$-magnetic specific heat and susceptibility. 
These features are related to a rather small Kondo energy scale of a few Kelvin in combination with a 
quasi-quartet crystal field ground state. 
Tuning the system towards long range magnetic order is accomplished by replacing a few at.\% of Ni 
by Cu or Co. Specific heat, susceptibility and resistivity studies reveal $T_{\rm N}\sim 0.2$\,K for 
CeNi$_8$CuGe$_4$ and $T_{\rm N}\sim 1$\,K for CeNi$_8$CoGe$_4$. 
To gain insight whether the transition from the paramagnetic NFL state to the magnetically ordered ground 
state is connected with a heavy fermion quantum critical point we performed specific heat and ac 
susceptibility studies and utilized the $\mu$SR technique and quasi-elastic neutron 
scattering.~\footnote{This workshop was supported in part by the Grant-in-Aid for the 
Global COE Program {\sl The Next Generation of Physics, Spun from Universality and Emergence} 
from the Ministry of Education, Culture, Sports, Science and Technology (MEXT) of Japan.}
\end{abstract}

\section{Introduction}

Some exciting novel electronic phenomena in solids, e.g.\ high-temperature 
and heavy-Fermion superconductivity, have been observed when tuning strongly correlated 
electron systems by substitution doping, pressure or other non-thermal parameters 
from a symmetry breaking ground state towards a symmetry conserving one,  e.g.\ 
from long range magnetic order towards a paramagnetic Fermi liquid ground state.
Prominent examples in this context are pressure induced, magnetically mediated 
unconventional superconducting states in antiferromagnetic (AF) Kondo lattice systems 
CePd$_2$Si$_2$, CeIn$_3$ \cite{mathur} and in ferromagnetic UGe$_2$ \cite{saxena}.   
If continuous, such symmetry breaking phase transition at virtually zero temperature
gives rise to quantum critical behavior at finite temperature (for review see e.g.~\cite{Sachdev})
which e.g.\ manifests in the non-Fermi liquid (NFL) behavior of various heavy Fermion 
systems~\cite{Stewart:2001,Loehneysen:2007}. 

CeNi$_9$Ge$_4$ is a very interesting paramagnetic Kondo lattice system that 
exhibits remarkable features such as a record value of the electronic specific heat 
coefficient with a paramagnetic ground state, 
$\gamma=C/T\simeq 5.5$ J/mol\,K$^2$ at 80\,mK~\cite{Michor:2004}, which is the highest among the 
strongly correlated electron systems.
CeNi$_9$Ge$_4$ further exhibits a strongly temperature-dependent Sommerfeld--Wilson ratio, 
$R=\chi/\gamma$, below 1 K and an approximate scaling of the $4f$-magnetic specific heat and 
susceptibility for the Ce concentration in a magnetically dilute solid 
solution Ce$_x$La$_{1-x}$Ni$_9$Ge$_4$~\cite{Killer:2004}.
Such scaling of the magnetic contributions to the specific heat and susceptibility 
reveals that the single ion type interactions, such as crystal field (CF) and Kondo effects,  
are responsible for the physical properties of these compounds.  
The CF ground state of  CeNi$_9$Ge$_4$ has been determined by means of single crystal susceptibility 
measurements which were analyzed in terms of the tetragonal crystal field model in combination 
with a Kondo screening correction by the poor man's scaling approach yielding a CF level scheme
with a  ground state formed by two doublets ($\Gamma_7^{(1)}$, $\Gamma_7^{(2)}$)  
split by about 0.5\,meV  and a well separated $\Gamma_6$ doublet with an excitation energy 
of 11\,meV~\cite{Michor:2006}.   
The low-temperature Kondo energy scale referring to the quasi-quartet ground state 
which results from the analysis of the macroscopic static susceptibility as well as 
microscopic neutron studies of the quasi-elastic linewidth is about 
0.3\,meV~\cite{Michor:2006}. 
The splitting of the $\Gamma_7^{(1)}$ and $\Gamma_7^{(2)}$ doublets is, thus, of comparable magnitude 
as the Kondo energy scale and thereby smeared to a quasi-quartet ground state which shows up in the 
specific heat as roughly $C\sim -T\ln T$, which is a typical behaviour observed in NFL systems 
at low temperature~\cite{Michor:2004,Killer:2004}.

Based on the above estimates of CF and Kondo energy scales of CeNi$_9$Ge$_4$ numerical renormalization
group calculations using the $SU(4)$ Anderson impurity model accounted for 
the basic thermodynamic features of CeNi$_9$Ge$_4$ reasonably well and demonstrated that the 
temperature dependent Sommerfeld--Wilson ratio $R(T)=\chi(T)/\gamma(T)$ results from an 
$SU(2)$ to $SU(4)$ cross-over~\cite{Anders:2006,Scheidt:2006}. 

\begin{figure}[t]
\centering
\includegraphics[width=0.90\textwidth]{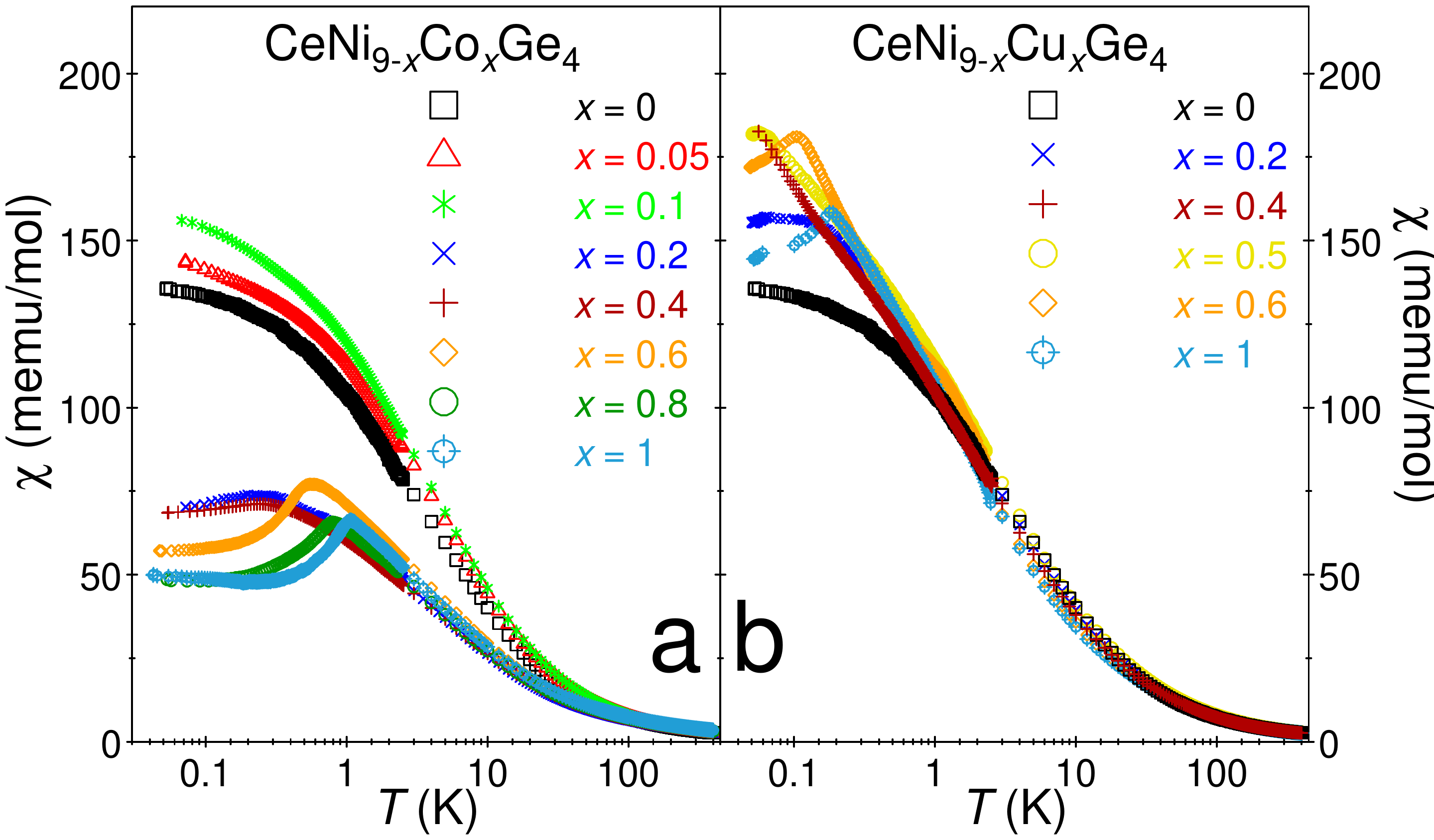}
\caption{\label{fig1} The magnetic susceptibility $\chi(T)$ of CeNi$_{9-x}$Co$_x$Ge$_4$ (a) 
and CeNi$_{9-x}$Cu$_x$Ge$_4$ (b) in semi-logarithmic plots (redrawn from~\cite{Peyker:2009,Peyker:2010}). 
AFM transitions are evident for $x>0.5$.}
\end{figure}

Cerium Kondo lattice systems having Kondo temperatures $T_\mathrm K$ as low as a few Kelvin usually 
exhibit a magnetically ordered ground state~\cite{sereni}, i.e.\ they are dominated by 
RKKY inter-site interactions.
CeNi$_9$Ge$_4$ with $T_\mathrm K\sim 3.5$\,K and no magnetic frustration
(tetragonal structure with easy direction of magnetisation in $c$-axis), 
however, displays a paramagnetic ground state. 
A factor in favor of a paramagnetic ground state is the effectively quasi-fourfold ground state 
degeneracy as proposed in the theoretical model of Coleman~\cite{Coleman:1983}. 
In this model above a critical value of the Kondo coupling constant, $(J\rho)_c$, the spin-compensated 
Kondo-lattice ground state is stable and this value $(J\rho)_c$ is shown to tend to zero as $O(1/N)$, 
providing new justification of applicability of the Kondo lattice model for rare earth based strongly 
correlated electron systems~\cite{Coleman:1983}.
To explore the significance of RKKY interactions in CeNi$_9$Ge$_4$ and to check for the possibility
of quantum criticality in this system, we have studied Ni-site substitutions 
by Cu and Co, i.e.\  equivalent to electron and hole doping, respectively. 
The aim to tune the system towards long range magnetic order is in fact accomplished by replacing a few 
at.\%  of Ni by Co as well as by Cu (see figure~1 and references~\cite{Peyker:2009,Peyker:2010} for 
further details including specific heat and resistivity studies). 
It is surprising to find magnetic order for both, CeNi$_8$CoGe$_4$ and CeNi$_8$CuGe$_4$, because 
electron and hole doping are expected to drive the Kondo temperature in opposite directions.
The latter is in fact supported by the susceptibility data displayed in figure~1 
where Co substitution in 1a clearly reduces the susceptibility~\cite{Peyker:2010}, 
i.e.\ indicating an increase of $T_{\rm K}$,
whereas Cu substitution in 1b tends to increase the low temperature susceptibility, thus, pointing
towards a reduction of $T_{\rm K}$ in the solid solution CeNi$_{9-x}$Cu$_x$Ge$_4$~\cite{Peyker:2009}.  
What both solid solutions, CeNi$_{9-x}T_x$Ge$_4$ with $T=$~Co and Cu, show in common is 
a reduction of the effectively quasi-fourfold degeneracy of the CF ground state of the parent CeNi$_9$Ge$_4$ 
compound towards a two-fold degenerate one in both CeNi$_8$CoGe$_4$ and CeNi$_8$CuGe$_4$ which is proposed
by the analysis of magnetic entropy data~\cite{Peyker:2009,Peyker:2010}.  

In this paper we present new microscopic studies of magnetic correlations in 
CeNi$_{9-x}$Cu$_x$Ge$_4$
by means of quasi-elastic neutron scattering and muon spin relaxation ($\mu $SR) experiments.  
Macroscopic studies of thermodynamic properties such as magnetic susceptibility, specific heat and
thermal expansion revealed a quantum critical point in this solid solution series where
CeNi$_{8.6}$Cu$_{0.4}$Ge$_4$ exhibits quantum critical behavior with $\chi$ and $C/T\propto -\ln T$ 
as well as a strongly composition dependent behavior of the Gr\"uneisen ratio 
$\mathnormal{\Gamma}(T)\propto\alpha(T)/C(T)$ at low 
temperatures where $\mathnormal{\Gamma}(T\rightarrow 0)$
increases by orders of magnitudes right at the critical composition 
CeNi$_{8.6}$Cu$_{0.4}$Ge$_4$~\cite{Peyker:2009}. 

\section{Experimental details}

Polycrystalline samples used for neutron scattering and $\mu $SR studies 
namely CeNi$_9$Ge$_4$, CeNi$_{8.6}$Cu$_{0.4}$Ge$_4$, CeNi$_{8}$CuGe$_4$, Ce$_{0.8}$La$_{0.2}$Ni$_9$Ge$_4$,
and LaNi$_{9}$Ge$_{4}$ were synthesized by high frequency induction melting on a water cooled 
copper hearth under a protective argon atmosphere. 
The starting materials were Ce and La metals (Ames MPC~\cite{ames}, 99.95\%), 
Ni (Johnson-Matthey, GB, 99.999\%) and zone-refined Ge ingots (Johnson-Matthey, GB, 99.9999\%).
In a first step, Ni and Ge were melted together to produce a master alloy, which was then alloyed with Ce metal.
The samples were finally annealed for one week at 1270\,K in evacuated quartz ampules.
Standard x-ray diffraction performed on powder revealed essentially
phase pure samples crystallizing in the tetragonal space group $I4/mcm$
where substitution of Ni by Cu leads to a linear volume increase 
reaching 0.8\%\ for CeNi$_{8}$CuGe$_4$ as compared to CeNi$_{9}$Ge$_4$.
Details of the structural characterization were reported earlier~\cite{Peyker:2009}. 

Cold neutron quasi-elastic scattering experiments were carried out
on the IN6 time-of-flight spectrometer at ILL Grenoble.
The spectrometer was operated with an incident neutron energy of
3.15\,meV (wavelength $\lambda=5.12$\,\AA), yielding an energy 
resolution of 70\,$\mu$eV at full width half maximum (FWHM).
Powder samples, each with masses near 20\,g, were enclosed in flat Al-can sample holders
yielding an effective sample thickness of about 1.3\,mm which was taken into 
account for proper absorption corrections.
For calibration we measured a vanadium reference sample of the same disc-shape 
geometry at 2\,K.  For subtracting the background signal of the sample holder 
and intrinsic non-magnetic contributions to the total scattering (mainly phononic ones), we
measured LaNi$_9$Ge$_4$ reference sample at same experimental conditions and
temperatures.   

The muon spin relaxation ($\mu$SR) experiments at temperatures down to 35\,mK
were performed on the {\sl MuSR} spectrometer at the ISIS facility where pulses of muons are 
implanted into the sample at 50\,Hz and with a FWHM of 70 ns. 
These muons are thermalized in the bulk of the sample within few ps and 
decay with a half-life, $\tau_{\mu}=2.2$\,$\mu$s into
positrons, which are emitted preferentially in the direction of the 
muon spin axis. Each positron is time stamped and therefore the muon spin
polarisation which corresponds to the asymmetry between positron counts in the 
forward and backward detectors (in so-called longitudinal geometry) can be followed 
as a function of time. In the present experiments, an initial asymmetry of 0.28 
refers to an initially perfect spin polarization of the implanted muons parallel to 
the beam forward direction.

\section{Cold neutron quasi-elastic studies}
\begin{figure}[t]
\begin{center}
\includegraphics[width=0.65\textwidth]{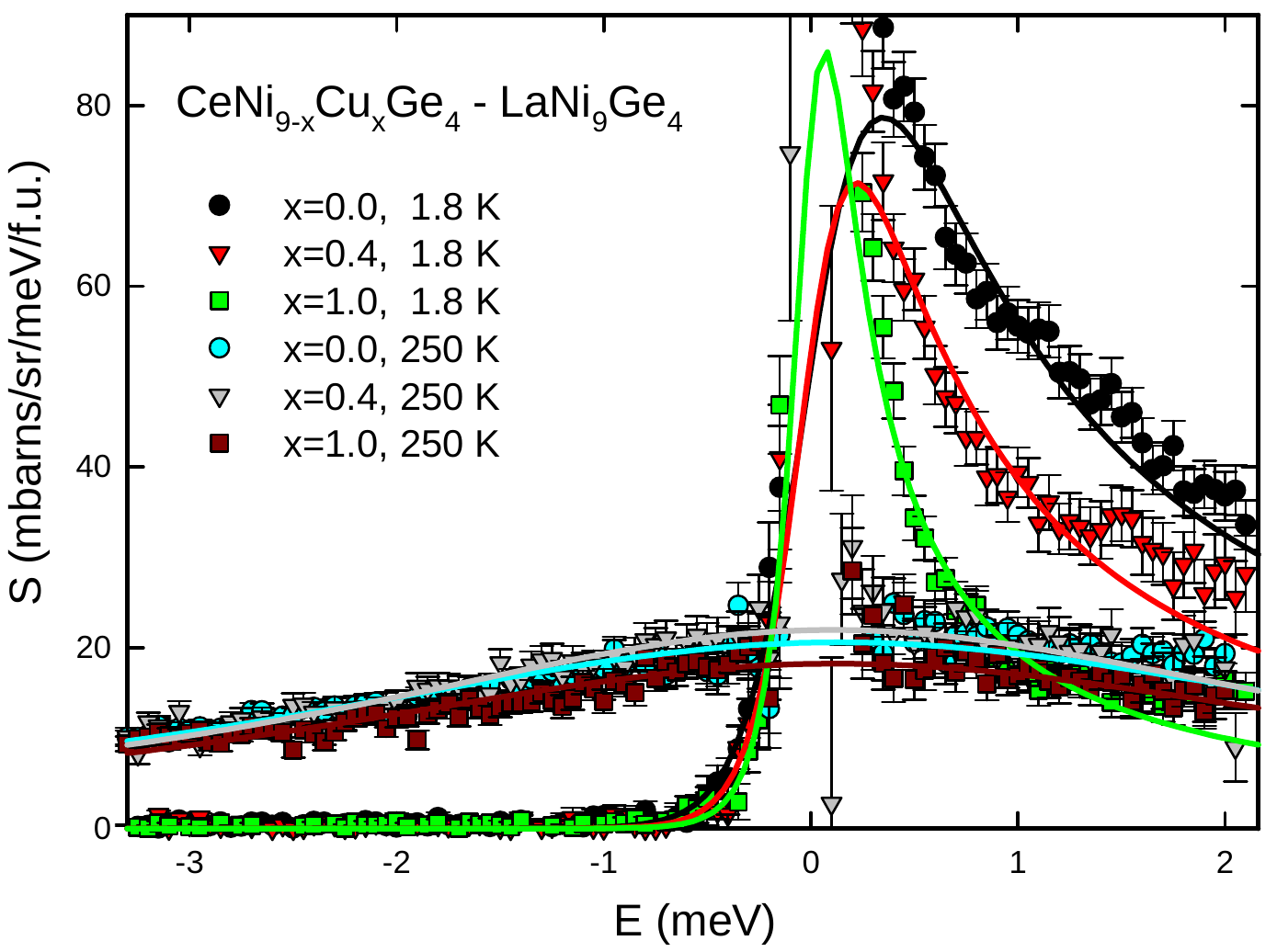}
\end{center}
\caption{\label{fig2} The magnetic correlation function $S_m(\omega)$ at 1.8\,K and 250\,K
for compositions CeNi$_{9-x}$Cu$_x$Ge$_4$ as labeled; solid lines are Lorentzian fits 
according to equation~\ref{Lor}.}
\end{figure}
The variation of the Kondo temperature in the solid solution CeNi$_{9-x}$Cu$_x$Ge$_4$
is evaluated from the quasi-elastic paramagnetic response probed by cold neutron scattering, i.e.\
via measurements of correlation functions $S(q,\omega)$ as described earlier in reference~\cite{Michor:2006}. 
As the coherent and incoherent cross sections of  CeNi$_{9-x}$Cu$_x$Ge$_{4}$ with $x\leq 1$ and LaNi$_{9}$Ge$_{4}$ 
match each other within a few percent, the magnetic scattering $S_m(q,\omega)$ of cerium 
is obtained by subtracting the LaNi$_{9}$Ge$_{4}$ data representing phonon plus the background contributions.
The magnetic scattering data $S_m(q,\omega,T\geq 1.8{\rm K})$ obtained thereby exhibit only a weak 
$q$-dependence of the intensity which corresponds well to the usual Ce$^{3+}$ form factor.
Accordingly,  
$q$-integrated (section from 0.4 to 1.0\,\AA$^{-1}$) magnetic correlation functions $S_m(\omega)$ of 
CeNi$_9$Ge$_4$, CeNi$_{8.6}$Cu$_{0.4}$Ge$_4$, and CeNi$_{8}$CuGe$_4$ measured at 1.8\,K and 250\,K 
are displayed in figure~2. 
The magnetic scattering,  
\begin{equation} \label{somega}
S_m(\omega,T)=\frac{\chi^{\prime\prime}(\omega,T)}{1-\exp(-\hbar\omega/k_{B}T)}, 
\end{equation} 
is related to the 
absorptive component $\chi^{\prime\prime}(\omega,T)$ of the dynamic susceptibility which
in case of a normal paramagnetic response with a single exponential decay of the magnetisation 
density is given by a Lorentzian
\begin{equation} \label{Lor}
\chi^{\prime\prime}(\omega,T)=\chi^{\prime}(T)\frac{\omega\Gamma(T)}{\omega^2+\Gamma^2(T)}
\end{equation} 
where $\Gamma (T)$ and $\chi^{\prime}(T)$ are the temperature dependent relaxation rate  
(line width) and static susceptibility, respectively. 
Despite of an underlying CF splitting, the quasi-elastic response is acceptably well 
described by simple Lorentzian fits according to equation~\ref{Lor}.
While all the high temperature data measured at 250\,K where $k_{\rm B}T$ is larger than the 
overall CF splitting are essentially on top of each other, there are rather significant changes of the
low temperature quasi-elastic response.

The Lorentzian fits indicated by solid lines in figure~2
reveal a reduction of the line width $\Gamma(1.8 {\rm K})$ from 0.47\,meV to 0.36\,meV and 0.20\,meV
for CeNi$_9$Ge$_4$, CeNi$_{8.6}$Cu$_{0.4}$Ge$_4$, and CeNi$_{8}$CuGe$_4$,
respectively. Simultaneously, also the overall integrated intensity which
corresponds to the static susceptibility $\chi^{\prime}(1.8 {\rm K})$ (see equation~\ref{Lor}) 
decreases from 0.104\,emu/mol to 0.088\,emu/mol, and 0.073\,emu/mol, respectively. 
Although the trend is similar the significant variation of $\chi^{\prime}(1.8 {\rm K})$ 
seems in contradiction with the SQUID susceptibility data in figure~1b 
showing a minor variation of $\chi(1.8 {\rm K})$.
A close agreement of the value proposed by the Lorentzian fit with the SQUID result is 
observed for CeNi$_9$Ge$_4$ (from CF parameters in reference~\cite{Michor:2006} 96\% of the total
magnetic response is expected from the $\Gamma_7^{(1)}$-$\Gamma_7^{(2)}$ quasi-quartet),
but an increasing lack of quasi-elastic intensity as compared to the static SQUID susceptibility
in the case of CeNi$_{8.6}$Cu$_{0.4}$Ge$_4$, and CeNi$_{8}$CuGe$_4$.
The latter is obviously connected with a growing splitting of the $\Gamma_7^{(1)}$ and $\Gamma_7^{(2)}$
doublets caused by Cu substitution which alters the local environment of the cerium ions and, thereby,
changes their CF scheme. 
The quasi-elastic neutron data, thus, reveal that Ni/Cu substitution causes a significant reduction of the 
Kondo energy scale and also a reduction of the effective ground state degeneracy towards a doublet in the
case of CeNi$_{8}$CuGe$_4$. The latter is also confirmed by inelastic neutron data 
measured recently with an incident neutron energy of $E_{\mathrm i}=20.5$\,meV at FRMII
where an approximate CF scheme with doublets at ground state, 5\,meV and 12\,meV is observed for 
CeNi$_{8}$CuGe$_4$~\cite{FRMII}.  

\section{Muon spin relaxation studies}

In order to probe magnetic correlations at low (dilution fridge) temperatures we 
utilized the $\mu $SR technique which is a local probe method sensitive to extremely small 
internal fields and ideally suited to detect any kind of static magnetism as well as dynamic 
magnetic features (if accessible in $\mu$s time scale of the muon life time). 
The $\mu $SR technique has, thus, been extensively applied to heavy fermion systems 
(see e.g.\ the review by Amato~\cite{amato}). 

\begin{figure}[t]
\begin{center}
\includegraphics[width=0.97\columnwidth]{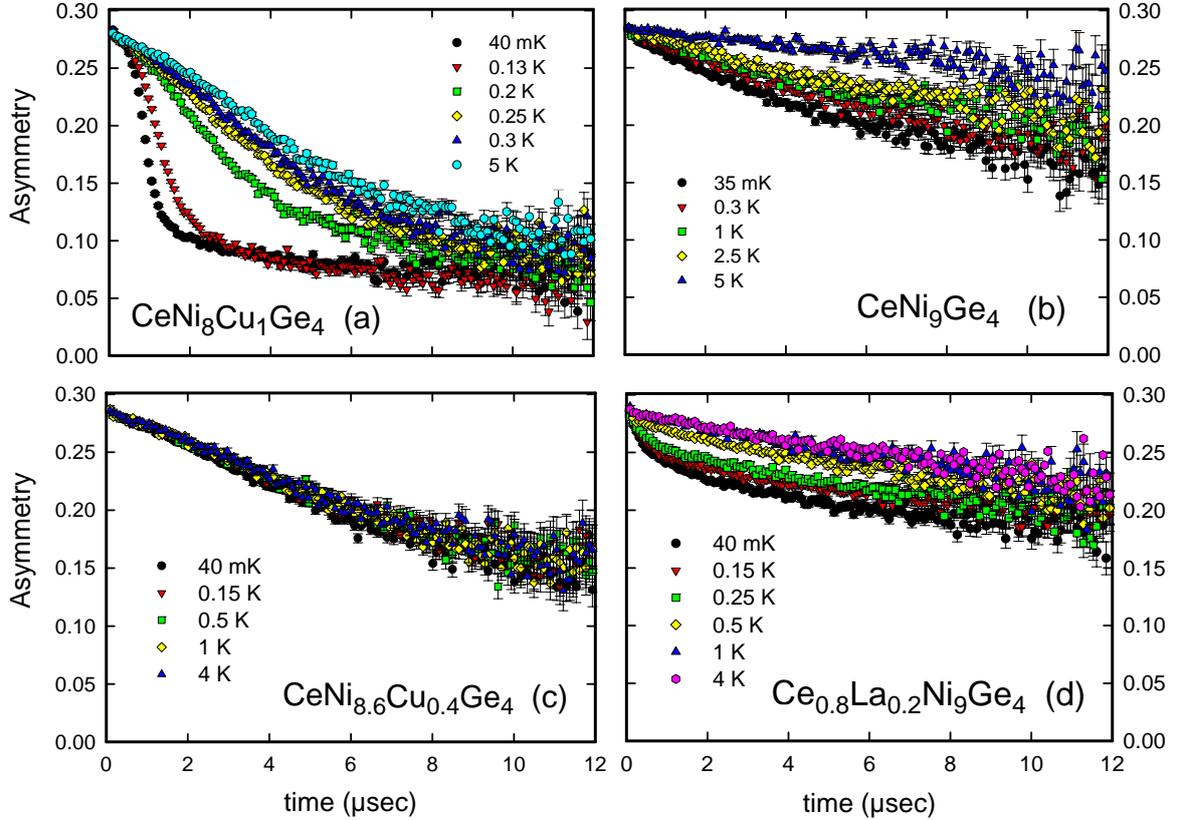}
\end{center}
\caption{The time dependent $\mu$SR asymmetry data of CeNi$_{8}$CuGe$_4$ (a), 
CeNi$_9$Ge$_4$ (b), CeNi$_{8.6}$Cu$_{0.4}$Ge$_4$ (c), and Ce$_{0.8}$La$_{0.2}$Ni$_9$Ge$_4$ (d)  
at selected temperatures and zero field.} 
\label{Fig3}
\end{figure}

The zero-field $\mu $SR measurements down to about 40\,mK for CeNi$_9$Ge$_4$,
CeNi$_{8.6}$Cu$_{0.4}$Ge$_4$, CeNi$_{8}$CuGe$_4$, and a magnetically dilute sample
Ce$_{0.8}$La$_{0.2}$Ni$_9$Ge$_4$ are displayed in figure~3 as time dependent
depolarization of the muon spins, i.e.\ asymmetry estimated from the positions 
detected in beam forward and beam backward directions.
All relaxation data measured at highest temperatures displayed in figure~3, i.e.\ at 4\,--\,5\,K,
are in approximate agreement with the Gaussian Kubo-Toyabe relaxation function 
(see e.g.\ reference~\cite{amato})
which refers a depolarisation of the muon spins by quasi-static nuclear dipolar fields. 
As the largest nuclear moments are contributed by copper atoms, CeNi$_{8}$CuGe$_4$
in figure~3a clearly displays the quickest relaxation at high temperatures and
nuclear dipolar fields maintain their dominant role down to 0.3\,K. 
Just near 200\,mK, where the susceptibility in figure~1b exhibits a sharp cusp  
indicating the onset of AF order, there is a dramatic change
of the $\mu$SR signal showing a rather quick depolarization towards 
a typical background asymmetry $A_{\rm bg}\sim 0.05$ at low temperatures. 
The absence of any sign of coherent frequency oscillations in the zero field $\mu$SR 
spectra which has been observed also in some other antiferromagnetic materials,
e.g.\ Ce$_8$Pd$_{24}$Ga~\cite{Adroja:2003}, refers to a muon stopping site
in a high symmetry position with respect to the Ce sublattice where dipolar fields 
of AF ordered Ce moments may compensate each other to zero.
In the presence of substitutional disorder in CeNi$_{8}$CuGe$_4$,
muon spins at such point of compensating dipolar fields should, of
course, experience some distribution of static internal fields with a zero mean 
at the muon site. 
This interpretation of the CeNi$_{8}$CuGe$_4$ 40\,mK data is supported by a markedly different 
field dependence of CeNi$_{8}$CuGe$_4$ observed in longitudinal field  
$\mu$SR measurements (not shown for the sake of brevity) as compared to 
all other compounds studied in this investigation.

The exponential low temperature relaxation of CeNi$_9$Ge$_4$ and magnetically dilute 
Ce$_{0.8}$La$_{0.2}$Ni$_9$Ge$_4$ in figure~3b and 3d refers a muon spin depolarization by 
dynamically fluctuating dipolar fields.
In the case of CeNi$_9$Ge$_4$ a clean simple exponential behaviour $A(t)\propto\exp{(- \lambda t)}$ 
with a depolarization rate $\lambda=0.133$\,$\mu$s$^{-1}$ at 35\,mK is observed (see figure~3b).
As already discussed in reference~\cite{Michor:2006} a simple exponential $\mu $SR signal rules 
out a disorder or inhomogeneity dominated NFL regime which would be indicated by a broad 
distribution and a strong temperature dependence of $\mu $SR relaxation rates~\cite{maclaughlin}.
The latter situation seems to apply for the solid solution Ce$_{0.8}$La$_{0.2}$Ni$_9$Ge$_4$ 
which exhibits a significantly larger initial relaxation rate and strongly stretched
exponential behaviour. The larger initial relaxation rate is counter-intuitive for the 
magnetically dilute case in particular when keeping in mind the approximate
concentration scaling of magnetic contributions to the specific heat and susceptibility
reported by Killer {\sl et al.}~\cite{Killer:2004}. 
These effects of magnetic dilution are even more remarkable when considering 
the corresponding results of CeNi$_{8.6}$Cu$_{0.4}$Ge$_4$ which
according to our thermodynamic studies~\cite{Peyker:2009} is located right at 
the critical concentration for the onset of AF order and clearly exhibits quantum critical 
behavior with $\chi$ and $C/T\propto -\ln T$. 
Against all expectations $\mu $SR results in figure~3c do not reveal any temperature dependence of the 
CeNi$_{8.6}$Cu$_{0.4}$Ge$_4$ data even from 40\,mK to 4\,K where muon spins 
essentially probe nothing else than the nuclear dipolar fields of the copper cores.

The only plausible interpretation of these puzzling results are strong AF short range magnetic
correlations causing a compensation dipolar fields at the muon site which is most effective 
right at the quantum critical point, i.e.\ muon will not see any change with temperature
because nuclear dipolar fields remain dominant at all temperatures.  
Magnetic dilution through Ce/La substitution of course
creates at least two kinds of muon stopping sites: (i) in between a pair of Ce-ions where
correlated $4f$ moments may compensate each other and (ii) in between a pair of Ce and La
where the muon spin feels the full Ce moment. Longitudinal field $\mu $SR data (not shown)
in fact support this scenario where a smaller number of muon sites ($\leq 40$\%) is of 
the second type and causes the quick initial relaxation and is more robust against the 
externally applied longitudinal field, i.e., connected with larger local fields at
the muon site, and a larger number of muon sites is of type (i) displaying a much 
slower relaxation which is easily suppressed by just a few 10\,G longitudinal field
similar to the field dependence seen in CeNi$_9$Ge$_4$ and CeNi$_{8.6}$Cu$_{0.4}$Ge$_4$.

\section{Summary}

The effectively quasi-fourfold ground state degeneracy in combination
with a Kondo temperature of only a few Kelvin places the paramagnetic  
heavy fermion system CeNi$_9$Ge$_4$ in an exceptional position 
among Kondo lattice systems where the paramagnetic ground state
could be stabilized by the quasi-quartet CF ground state.
The latter is suggested by the fact that tuning CeNi$_9$Ge$_4$ towards a magnetic 
ground state is accomplished by Ni/Cu as well as Ni/Co substitution
(equivalent to electron and hole doping, respectively) which are both leading to
a slightly modified local environment of the cerium ions and, thereby,
cause a reduction of the CF ground state towards a doublet one.
These modifications of ground state degrees of freedom are 
revealed by microscopic studies via cold neutron quasi-elastic scattering on CeNi$_9$Ge$_4$ 
and the solid solutions CeNi$_{8.6}$Cu$_{0.4}$Ge$_4$ and CeNi$_{8}$CuGe$_4$
where a significant narrowing of the quasi-elastic line at low
temperatures and simultaneously some loss in quasi-elastic intensity is observed.
The latter indicates a transfer of quasi-elastic to inelastic scattering
caused by a splitting of the quasi-quartet CF ground state towards two well 
separated doublets.

Local probe $\mu$SR studies performed down to temperatures near 40\,mK
revealed rather unexpected results where on the one hand, 
magnetic dilution (Ce/La substitution) tends to increase the local fields experienced 
by the implanted muons, 
whereas on the other hand, tuning towards quantum criticality in CeNi$_{8.6}$Cu$_{0.4}$Ge$_4$  
tends to annihilate the dipolar fields of cerium moments at the muon stopping sites.
A cancellation of the dipolar fields of cerium $4f$ moments is only possible if  
muons locate in a high symmetry position with respect to the Ce sublattice 
where strong AF correlations lead to a pairwise compensation of cerium dipolar fields.
The $\mu$SR results are, thus, suggestive of strong AF short range dynamic correlations being
present in the unusual Kondo lattice ground state of CeNi$_9$Ge$_4$.

\ack
This research project was supported by COST P-16 ECOM. 
The $\mu$SR studies at the ISIS facility were supported by the European Commission under 
the $7^{\rm th}$ Framework Programme through the Key Action: 
Strengthening the European Research Area, Research Infrastructures
(Contract CP-CSA-INFRA-2008-1.1.1 Number 226507-NMI3).

The authors thank the Yukawa Institute for Theoretical Physics at Kyoto University. 
Discussions during the YITP workshop YITP-W-10-12 on 
"International and Interdisciplinary Workshop on Novel Phenomena 
in Integrated Complex Sciences: from Non-living to Living Systems" 
were useful to complete this work.

\end{document}